\documentclass{article}
\pdfoutput=1

\usepackage[preprint,nonatbib]{nips_2018}

\usepackage{microtype}
\usepackage{graphicx}
\usepackage{subfigure}
\usepackage{booktabs} 
\usepackage{amsfonts} 
\usepackage{graphicx}

\usepackage[utf8]{inputenc} 
\usepackage[T1]{fontenc}    
\usepackage{hyperref}       
\usepackage{url}            
\usepackage{booktabs}       
\usepackage{amsfonts}       
\usepackage{nicefrac}       
\usepackage{microtype}      
\usepackage{amsmath}
\usepackage{listings}

\title{Fashionable Modelling with Flux}

\author{
  Michael J Innes \\
  Julia Computing, Inc. \\
  Edinburgh, UK \\
  \texttt{mike.j.innes@gmail.com} \\
  \And
  Elliot Saba \\
  Julia Computing, Inc. \\
  Cambridge, MA, USA \\
  \texttt{staticfloat@gmail.com} \\
  \And
  Keno Fischer \\
  Julia Computing, Inc. \\
  Cambridge, MA, USA \\
  \texttt{keno@juliacomputing.com} \\
  \And
  Dhairya Gandhi \\
  Julia Computing, Inc. \\
  Bangalore, India \\
  \texttt{dhairya@juliacomputing.com} \\
  \And
  Marco Concetto Rudilosso \\
  University College London \\
  London, UK \\
  \texttt{marco.rudilosso.15@ucl.ac.uk} \\
  \And
  Neethu Mariya Joy \\
  Birla Institute of Technology and Science \\
  Pilani, India \\
  \texttt{neethumariyajoy@gmail.com} \\
  \And
  Tejan Karmali \\
  National Institute of Technology \\
  Goa, India \\
  \texttt{tejank10@gmail.com} \\
  \And
  Avik Pal \\
  Indian Institute of Technology \\
  Kanpur, India \\
  \texttt{avikpal@cse.iitk.ac.in} \\
  \And
  Viral B. Shah \\
  Julia Computing, Inc. \\
  Cambridge, MA, USA \\
  \texttt{viral@juliacomputing.com} \\
}

\begin{document}

\maketitle

\begin{abstract}
Machine learning as a discipline has seen an incredible surge of interest in recent years due in large part to a perfect storm of new theory, superior tooling, renewed interest in its capabilities.  We present in this paper a framework named \texttt{Flux} that shows how further refinement of the core ideas of machine learning, built upon the foundation of the Julia programming language, can yield an environment that is simple, easily modifiable, and performant.  We detail the fundamental principles of \texttt{Flux} as a framework for differentiable programming, give examples of models that are implemented within \texttt{Flux} to display many of the language and framework-level features that contribute to its ease of use and high productivity, display internal compiler techniques used to enable the acceleration and performance that lies at the heart of \texttt{Flux}, and finally give an overview of the larger ecosystem that \texttt{Flux} fits inside of.
\end{abstract}

\section{Introduction}

\texttt{Flux} is a new machine learning (ML) stack. Only a year old and developed by a very small team, it nonetheless has garnered an enthusiastic and rapidly growing community of researchers and industry users, building new kinds of models and techniques.

ML engineering is fundamentally a programming languages problem; machine learning models are growing ever more complex, incorporating control flow, state and data structures, and borrowing techniques from research areas throughout the field of computer science from software engineering to scientific simulation to statistical inference.  These models are best viewed as \textit{differentiable algorithms}; and to express algorithms we use programming languages \cite{innes2018machine}.

\texttt{Flux} takes Julia \cite{bezanson2017julia} to be this language. Julia is recent but, being designed from the ground up for mathematical and numerical computing, unusually well-suited for expressing ML programs.  Its mix of modern design and novel techniques in the compiler---which we have significantly extended to further suit differentiable algorithms and accelerators---makes it easier to address the high performance requirements needed for applying machine learning to large models and data sets.

There are three pillars that set \texttt{Flux} apart among ML systems: simplicity, hackability, and underlying compiler technology.

\subsection{Simplicity}

The word ``simple" is typically not thrown around in the ML systems world. However, simplifying solutions is a crucial part of approaching more complex problems. Where typical ML frameworks are written in many hundreds of thousands of lines of C++ \cite{abadi2016tensorflow}, \texttt{Flux} is only a thousand lines of straightforward Julia code.

Several factors compound to enable this. First, the ability to write everything from layer definitions, to algorithmic differentiation, to CUDA kernels, in high-level and \textit{fast} Julia code enables high developer productivity; code can typically be written once and forgotten about rather than being written, rewritten and optimized in C over years. Fast high-level code also enables something much more powerful than convenience: abstraction. This means infrastructure like higher-order kernels for \texttt{map(f,xs)} and \texttt{broadcast}, which can be written once and generalize to any user-defined \texttt{f} or input type with no extra effort or undue performance overhead.

This extends to integration with other tools in the ecosystem.  Writing an image pipeline really means loading the \texttt{Flux} and \texttt{Images} \cite{Images.jl} packages, and using them as normal. In fact, even GPU support is handled this way; rather than being provided as part of \texttt{Flux}, one loads a generic GPU arrays package such as \texttt{CuArrays} \cite{CuArrays.jl}, and passing those arrays into a \texttt{Flux} model just works; transferring data and computational kernels off to the GPU accelerator then bringing the results back to the CPU without any special handling within \texttt{Flux} itself. This leverages Julia's heavy use of specialization, which effectively generates custom code for the GPU, and even for custom floating-point types that would otherwise need to be written by hand.  Far from becoming an all-encompassing monolith, \texttt{Flux} remains a lean ``glue'' package, bringing together a set of underlying abstractions (e.g. gradients and GPU support) and combining them to create an ML framework.

\subsection{Hackability}

A core tenet of \texttt{Flux}, and Julia more generally, is that library code is just user code that happens to have been loaded from an external file. Unlike previous ML frameworks (including those written in Julia), \texttt{Flux} does not pick a certain level of abstraction (such as mathematical graphs or layer stacking) that all models must use. Instead, careful design of the underlying automatic differentiation (Section \ref{zygote}) allows freely mixing mathematical expressions, built-in and custom layers and algorithms with control flow in one model. This makes \texttt{Flux} unusually easy to extend to new problems.

\texttt{Flux} users regularly inject custom CUDA kernels, write down new mathematical functions and layers, hook in custom gradient definitions and even custom parallel training algorithms, all with the same performance as built-in features.  Hooking in custom functionality is often a one-line change that happens in the same Julia script as model code.  Because there is no difference between \texttt{Flux} code and other general Julia code, it is possible to break out of the typical deep learning paradigm and experiment with new concepts such as interleaving gradient descent training with MCMC.  Just write down your own training loop; the obvious mathematical code will work, and be fast.

This extends to integrating other packages with \texttt{Flux} models, such as Julia's state-of-the-art tools for mathematical optimization \cite{dunning2017jump}, differential equations \cite{rackauckas2017differentialequations} and probabilistic programming \cite{turing18}. This goes deeper than just using neural nets alongside these other techniques, as one can even incorporate them directly into models. For example, Julia's differential equation solvers---while not explicitly adapted either for AD or GPUs---can seamlessly be used with both. This means that one can use a physical simulation to enhance the predictions of the model, then backpropagate through the entire model \textit{including} the simulation to get gradients.  Bringing tools like physics simulators into models is where deep learning truly becomes differentiable programming.

\subsection{Compiler Technology}

\texttt{Flux} is committed to providing a dynamic (or ``define-by-run'') interface, and takes a hard line against any kind of graph building or performance annotations \cite{PyTorch1}. We support all of Julia's language features, from control flow and data structures to macros. Users can code interactively in Jupyter notebooks and combine high-performance numerics with convenient plotting and visualization. But we also want to get the benefits traditionally held by ``static graph" frameworks---zero-overhead source-to-source AD, operator fusion, multi-GPU/distributed training, and single-executable deployment.

Doing this effectively requires extracting and analyzing ``static graphs'' directly from written Julia syntax; but this really just describes the task of a \textit{compiler}. Most ML systems problems are, in fact, standard and well-studied compiler problems, viewed through the right lens. Using a compiled language is enough to solve many issues, and extending that compiler is the best way to solve many more.

Several illustrative examples, covering differentiation (Section \ref{zygote}), GPU and TPU compilation (Sections \ref{gpu} and \ref{tpu}) and SPMD / batching (Section \ref{spmd}) are covered briefly in this paper.

\section{Fashionable Modelling}

\subsection{Discriminative Adversarial Networks}

Nonstandard gradient control is a growing need in the machine learning community as model architectures are stretched to complete more and more complex tasks.  A significant recent development has been the emergence of adversarial networks \cite{GANs}, where two networks are arranged in opposition to each other, one attempting to learn a task and the other inventing new inputs that the first can learn from.  Extensions of this idea include work done in \cite{Saba2018} using \texttt{Flux}, where adversarial networks were used to decrease bias induced by dataset imbalance.

The fundamental problem was strong correlation between classification label (in this case, \textit{tuberculosis} versus \textit{non-tuberculosis} coughs) and originating dataset (in this case, \textit{clinical} versus \textit{non-clinical}).  The machine learning model naturally learned to key off of the differences in dataset, rather than the differences in the cough sounds themselves, because the differences between clinical and non-clinical recordings were larger than the differences between tuberculosis and non-tuberculosis coughs within a single dataset.  Due to the fact that the overwhelming majority of non-tuberculosis coughs were from the non-clinical dataset, the classification algorithm was trapped in a local minimum simply predicting ``tuberculosis'' for every sample from the clinical dataset.

\begin{figure*}[ht]
  \centering
  \includegraphics[width=.95\linewidth]{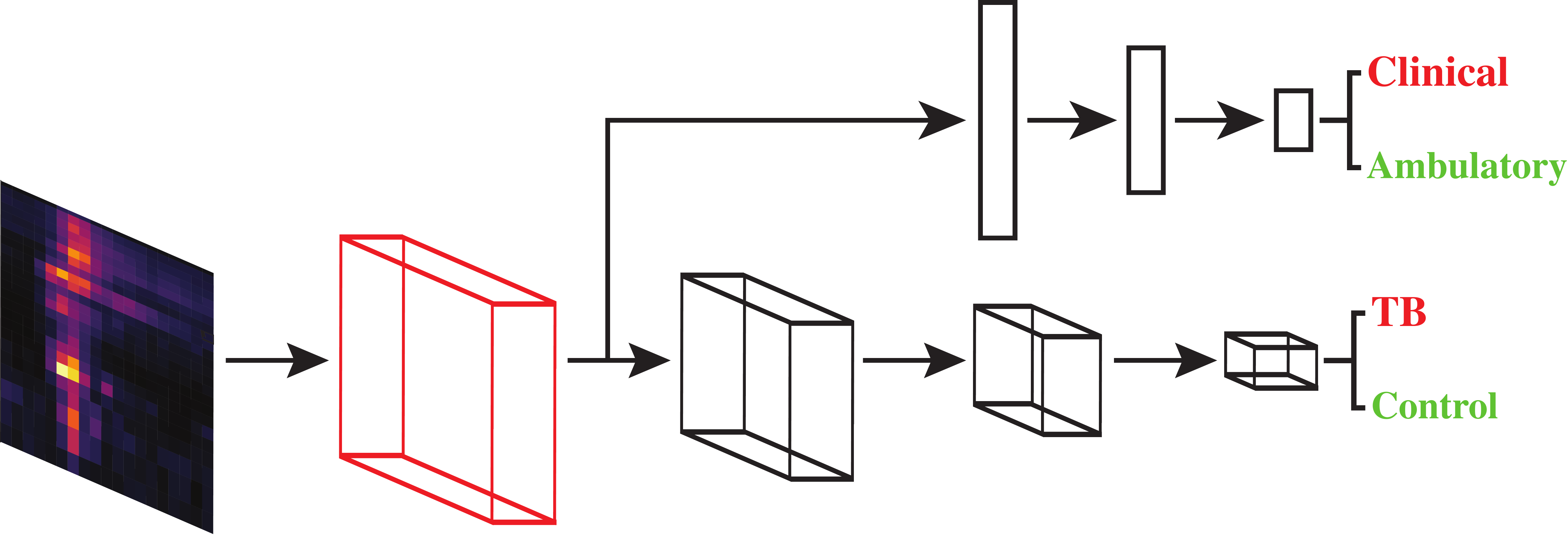}
  \caption{Cough classification architecture with DAN and classifier architectures visualized.  Along the bottom lies the convolutional classification network that classifies cough type, with the outputs from the first layer feeding into a multilayer perceptron that classifies dataset source.}
  \label{fig:dirty_dan}
\end{figure*}

In order to solve this, a Discriminative Adversarial Network (DAN) was employed, building a second network explicitly designed to determine dataset provenance.  This network is then used to ``penalize'' a shared set of weights (the convolutional block shown in red in Figure~\ref{fig:dirty_dan}) for extracting information that can be used to determine which dataset a sample originated from.

\begin{figure}[ht]
\begin{verbatim}
for x, y_c, y_d in training_set
    y_c_hat, y_d_hat = model(x)

    c_loss = loss(y_c_hat, y) +
             lambda*loss(y_d_hat, 1 - y_d)
    d_loss = loss(y_d_hat, y_d)

    back!(c_loss)
    back!(d_loss)

    opt()
end
\end{verbatim}
\caption{DAN training loop}
\label{listing:dirty_dan_training_loop}
\end{figure}

The code necessary to perform this nonstandard optimization task is surprisingly simple; shown in Figure~\ref{listing:dirty_dan_training_loop}, it simple calculates a forward pass through the model, returning the outputs from the two branches of the model as \texttt{y\_hat} and \texttt{y\_dan\_hat}, calculates the classifier loss (\texttt{c\_loss}) and the DAN loss (\texttt{d\_loss}), backpropagates the losses onto the network, then takes an optimizer step (\texttt{opt()}).  As can be seen, the optimization framework of \texttt{Flux} follows very naturally from the basic mathematical principles of optimization and provides for very flexible training and evaluation paradigms.

\section{Extending Julia's Compiler}

\subsection{Compiling Julia for GPUs}
\label{gpu}

Julia supports the basic CUDA programming model for writing GPU kernels \cite{besard2017effective}. A simple vector addition kernel looks similar to the CUDA C equivalent (Figure~\ref{listing:basic_cuda}).

\begin{figure}[h]
\begin{verbatim}
function kernel_vadd(a, b, c)
    i = (blockIdx().x-1) * blockDim().x +
        threadIdx().x
    c[i] = a[i] + b[i]
end
\end{verbatim}
\caption{Basic CUDA programming model expressed in Julia}
\label{listing:basic_cuda}
\end{figure}

However, Julia's type specialization enables a powerful set of additional abstractions on the GPU. For example, the code above is not restricted to dense arrays of floats, and could instead be given sparse arrays of complex numbers; Julia's normal specialization mechanisms would generate a new set of PTX instructions for that case. We can even abstract this code further into a ``higher-order kernel" that accepts the \texttt{+} function (or \texttt{*}, or arbitrary user-defined \texttt{f}) and thus create a whole family of functions \texttt{map(f, x, y)} in four lines of code \cite{innesGPU}.

Since this works for user-defined types, we can additionally make use of dual numbers for forward-mode differentiation \cite{revels2016forward}. Compared to reverse mode techniques, dual numbers have the significant advantage of being stack-allocated and interleaving primal and tangent computation, and thus in memory bound situations (such as GPUs) the derivatives are effectively free. This is particularly valuable in functions that contain control flow.

In the case of element-wise operations, such as \texttt{c = tanh.(a .+ b)} with vectors of length $N$, we can see this single $\mathbb{R}^{2N} \rightarrow \mathbb{R}^N$ computation as $N$ independent $\mathbb{R}^2 \rightarrow \mathbb{R}$ ones, for which forward mode is very efficient. In other words, by running the same computation with dual numbers we efficiently compute the (very sparse) Jacobians $\partial c / \partial a$ and $\partial c / \partial b$ fused with the original operation, which can then be used within the reverse-mode sweep to get gradients. In the base case this allows us to fully fuse elementwise operations like the above, but it also generalizes to complex user-defined functions including control flow, and can be much faster than the equivalent series of vector-level operations that would have to be used otherwise.

\subsection{Algorithmic Differentiation}
\label{zygote}

Pushing the limits of reverse-mode differentiation \cite{speelpenning1980compiling}, we have also come to see this as a language-level problem. Differentiation is a symbolic transformation that works on programs or symbolic expressions, and this is the domain of compilers. Existing ML frameworks achieve this transformation by \textit{tracing} (or \textit{partial evaluation}); a new tensor type is introduced which records all the basic mathematical operations performed, yielding a graph with the control flow and data structures of the host language elided. This graph, equivalent to a Wengert list \cite{bartholomew2000automatic}, is more easily differentiated than the original program.

Frameworks are then divided into the ``dynamic'' and ``static'' approaches \cite{neubig2017dynet} depending on whether they interpret \cite{maclaurin2015autograd} or compile \cite{bergstra2011theano} the recorded graph. However, this presents to us a difficult tradeoff: we either accept the overhead of an interpreter or freeze user control flow and limit the kinds of models that can be built (perhaps providing limited additional primitives to place control flow into the graph).

We reject these constraints and assert that the ``graph'' can instead  simply be the Julia syntax tree.  Extending previous work on differentiable languages \cite{pearlmutter2008reverse} we have built \texttt{Zygote}, a source-to-source auto-differentiator in Julia, which works directly on SSA-form IR and supports language features like control flow, recursion, data structures and macros, resolving the AD trade-off. By putting the generated SSA-form adjoint code through a traditional compiler such as LLVM \cite{lattner2004llvm}, we get all the benefits of traditional compiler optimization applied to both our forward and backwards passes. In addition, it opens up the possibility of extending that compiler infrastructure with more advanced and domain-specific optimizations, such as kernel fusion and compilation to accelerators such as TPUs.

\subsection{Compiling Julia for TPUs}
\label{tpu}

Google recently introduced a new API that allows users of their Cloud TPU offering to directly generate IR for the \textit{XLA} (``Accelerated Linear Algebra'') compiler. This IR is essentially a general purpose IR and optimizing compiler for expressing arbitrary computations of linear algebra primitives and thus provides a good foundation for targeting TPUs by non-Tensorflow users as well as for non-machine learning workloads.
We take advantage of Julia's dynamic type system and extendable compiler to build the capability to compile arbitrary Julia code to XLA, and then to run it on TPUs.
This allows users to take advantage of the full expressiveness of the Julia programming language in writing their models, including multiple dispatch, higher order functions and existing libraries such as those for differential equation solvers and generic linear algebra routines, while reaping the benefits of the high-performance systolic array engine within the TPU.

The basic workflow of the Julia compiler is to analyze chunks of a Julia program, identify static sub-segments, compile those sub-segments, and run them, returning to a dynamic environment only when necessary to compute, at runtime, the next static sub-segment to be compiled/run.
By hooking into this compilation infrastructure, we are able to define custom compiler passes that identify static sub-segments of Julia code and compile them directly to blocks of static XLA IR, which can be run directly on TPUs through the newly exposed API offered by Google.

While many dynamic languages are capable of limited amounts of static analysis, it is worth pointing out that the ``static sub-segments'' being referred to here are often quite large due to the Julia compiler's aggressive type specialization, type inference and constant propagation.
As an illustrative example, we are able to compile the VGG19 \cite{vgg19} machine learning model's forward pass, backward pass and optimization step (that is to say, the entire training loop) into a single chunk of XLA code that can be run on the TPU without breaking out into Julia at all.
This technical capability is critical in a heterogeneous computing system such as the TPU, as communication latency between the Julia host process and the TPU accelerator is significant, and must be avoided whenever possible.

\subsection{Automatic Batching}
\label{spmd}

The naturally data-parallel structure of many ML models makes them an excellent fit for massively parallel processors such as GPUs \cite{oh2004gpu} and TPUs. To get the most from these accelerators---which can have significant constant overheads per kernel launch, but scale very well over input size---it is common to \textit{batch} them, applying the forwards and backwards passes to multiple training examples at once. In simple models this can be achieved by stacking a set of images or samples along an additional batch dimension, and primitives such as matrix multiply, convolution and broadcasting naturally handle this extra dimension as if all samples were independent.

However, this task becomes much harder when dealing with variably-structured inputs, such as trees \cite{treelstm} or graphs \cite{graphconv}, or any kind of value-dependent control flow. Most researchers address this by taking on the significant burden of batching code by hand. Different solutions have been proposed for different frameworks \cite{neubig2017fly} \cite{looks2017deep}, which heuristically try to batch some high level operations together when possible, using different policies to choose what to batch and when, but these have their own usability issues and typically do not achieve the performance of hand-written code.

We propose that this problem is identical to that of Single Program Multiple Data (SPMD) programming, which has been well-studied by the language and compiler community for decades \cite{blelloch1990vector, Allen:1983:CCD:567067.567085}. Indeed, it is very similar to the model of parallelism used by GPUs internally, and has been implemented as a compiler transform for the SIMD units of CPUs \cite{Pharr2012ispcAS}. Taking inspiration from this work, we are implementing the same transform \cite{hack2011whole} in Julia to provide SPMD programming both for scalar SIMD units and for model-level batching. This allows us to reach the ideal of writing simple code that operates on individual samples, while still getting the best performance on modern hardware.

\subsection{JavaScript Compilation}

In an approach similar to that of the GPU and TPU compilation efforts, the \texttt{FluxJS} package \cite{FluxJS.jl} supports compiling Julia models to Javascript, using an approach based on partial evaluation of model code.  By defining a small number of fundamental operator equivalencies between Julia and Javascript, this package is able to parse the Julia syntax tree to create a function call graph, including control flow such as in RNNs, and makes use of the \texttt{tensorflow.js} \cite{tensorflow.js} package in order to accelerate specific machine learning operations.

\section{Conclusion}

In conclusion, we have presented the reasons why \texttt{Flux} in particular and Julia in general provide an excellent environment for high performance, simple and hackable machine learning.  We gave examples of complex models and functions defined in clearly readable, mathematically recognizable forms and detailed many of the techniques used to ensure that the generated code is not only performant on traditional CPUs, but also the accelerators that are increasingly critical to applied machine learning today.



\small

\bibliography{nips_2018}
\bibliographystyle{abbrv}

\end{document}